\documentstyle[12pt,epsf]{article}
\newcommand{\be}{\begin{equation}}
\newcommand{\ee}{\end{equation}}
\newcommand{\bea}{\begin{eqnarray}}
\newcommand{\eea}{\end{eqnarray}}
\newcommand{\bdm}{\begin{displaymath}}
\newcommand{\edm}{\end{displaymath}}
\newcommand{\no}{\nonumber \\}

\newcommand{\Mbar}{\;\overline{\rule[0.75em]{0.8em}{0em}}\hspace{-1em}M}
\newcommand{\MI}{M_{\mbox{\scriptsize I}}}
\newcommand{\MIhat}{\hat{M}_{\mbox{\scriptsize I}}}
\newcommand{\I}{{\mbox{\scriptsize I}}}
\newcommand{\R}{{\scriptscriptstyle R}}
\renewcommand{\L}{{\scriptscriptstyle L}}
\newcommand{\Ism}{{\mbox{\tiny I}}}

\newcommand{\ubar}{\overline{\rule[0.42em]{0.4em}{0em}}\hspace{-0.5em}u}
\newcommand{\dbar}{\,\overline{\rule[0.65em]{0.4em}{0em}}\hspace{-0.6em}d}

\newcommand{\Kbar}{\,\overline{\rule[0.75em]{0.7em}{0em}}\hspace{-0.85em}K}

\newcommand{\sA}{s_{\hspace{-0.15em}A}}
\newcommand{\al}{&\!\!\!\!}
\newcommand{\fs}{\; \; .}
\newcommand{\co}{\; \; ,}

\begin{document}

\begin{titlepage}

\begin{flushright}
BUTP--95/2
\end{flushright}

\begin{center}
{\LARGE {\bf \rule{0em}{4em}Dispersive analysis of the decay  $\eta\rightarrow
3\pi$\rule{0em}{1em} }}\\
\vspace{0.8cm}
A. V. Anisovich\\Academy of Sciences, Petersburg Nuclear Physics Institute\\
Gatchina, St. Petersburg, 188350 Russia\\\vspace{0.2cm} and\\ \vspace{0.2cm}
H. Leutwyler\\Institut f\"{u}r theoretische Physik der Universit\"{a}t
Bern\\Sidlerstr. 5, CH-3012 Bern, Switzerland and\\
CERN, CH-1211 Geneva, Switzerland\\
\vspace{0.6cm}
January 1996\\
\vspace{0.6cm}
{\bf Abstract} \\
\vspace{1.2em}
\parbox{30em}{We demonstrate that the decay $\eta\rightarrow 3\pi$ represents a
sensitive probe for the breaking of chiral symmetry by the quark masses.
The transition amplitude is
proportional to the mass ratio $(m_d^2-m_u^2)/(m_s^2-\hat{m}^2)$.
The factor of proportionality is calculated by means of dispersion
relations, using chiral perturbation theory to determine the subtraction
constants. The theoretical uncertainties in the result
are shown to be remarkably
small, so that $\eta$-decay may be used to accurately measure this
ratio of quark masses.} \\ \vspace{1em}
\rule{30em}{.02em}\\
{\footnotesize Work
supported in part by Schweizerischer Nationalfonds}
\end{center}
\end{titlepage}

{\it 1. Chiral perturbation theory.} Since Bose statistics does not
allow three pions to form a configuration where the total angular momentum
and the total isospin both vanish, the decay $\eta\rightarrow  3\pi$
violates isospin symmetry. The
transition amplitude contains terms proportional to the quark mass
difference $m_d-m_u$ as well contributions $\propto e^2$, generated by the
electromagnetic interaction. The latter are
strongly suppressed by chiral symmetry	-- the transition is
due almost exclusively to the isospin breaking
part $\frac{1}{2}(m_u-m_d)(\ubar u-\dbar d)$ of the QCD
Hamiltonian \cite{Sutherland}.
Current algebra implies that the matrix element relevant for
the transition
$\eta\rightarrow \pi^+\pi^-\pi^0$ is given by \be\label{lett1} A =
-\epsilon_0\,\frac{1}{F_\pi^2}\,(s-\mbox{$\frac{4}{3}$}\,M_\pi^2)
\co\hspace{3em}
\epsilon_0\equiv \frac{\sqrt{3}}{\,4}\,\frac{(m_d-m_u)}{(m_s-\hat{m})}\co\ee
where $F_\pi=92.4\,\mbox{MeV}$ is the pion decay constant,
$s=(p_{\pi^+}+p_{\pi^-})^2$ and $\hat{m}$
denotes the mean mass of the $u$- and
$d$-quarks, $\hat{m}\equiv\frac{1}{2}(m_u+m_d)$.

The current algebra formula represents the leading contribution of the chiral
perturbation series, which describes the decay amplitude as a sequence of
terms involving increasing powers of momenta and quark masses. The corrections
of first nonleading order (chiral perturbation theory to one
loop) are also known \cite{GL eta}. If the amplitude
is written in the form \be\label{amplitude}
  A=-\frac{1}{Q^2}\,\frac{M_K^2}{M_\pi^2}\,\frac{M_K^2-M_\pi^2}
{3\sqrt{3}F_\pi^2}\,M(s,t,u)\co\;\;\;
\frac{1}{Q^2}\equiv\frac{m_d^2\!-\!m_u^2}{m_s^2\!-\!\hat{m}^2}\co\ee
the one loop result for the dimensionless factor $M(s,t,u)$ exclusively
involves
measured quantities. It is of the form
\be\label{isospin decomposition} M(s,t,u)= M_0(s)
+(s\!-\!u)M_1(t)+(s\!-\!t)M_1(u)
+M_2(t)+M_2(u)-\mbox{$\frac{2}{3}$}M_2(s)\co\ee
with $t=(p_{\pi^-}+p_{\pi^0})^2$,
$u=(p_{\pi^0}+p_{\pi^+})^2$. The functions $M_0(s),M_1(s),M_2(s)$
represent
contributions with isospin I = 0, 1, 2, respectively.
They contain
branch point singularities generated by the final state interaction in the S-
and P-waves. The imaginary parts due to partial waves with angular
momentum $\ell\geq 2$ only start showing up if the chiral perturbation
series is evaluated to three or more loops. For this reason, the
amplitude retains the structure (\ref{isospin decomposition}) also at
next-to-next-to-leading order.
We stick to this approximation throughout the following, i.e.
represent the amplitude in
terms of the S- and P-waves and neglect the
discontinuities from partial waves with $\ell\!\geq\!2$.\footnote{The
same approximation is commonly used also in $\pi\pi$ scattering, where
the Roy equations \cite{Roy} provide a rigorous starting point for the
dispersive analysis. The validity of the representation analogous to
(\ref{isospin decomposition}) to two loop order of chiral perturbation
theory was pointed out in ref.\cite{Stern 1993} and an analysis of the $\pi\pi$
data on this basis is given in ref.\cite{Stern 1995}. The explicit two-loop
result
\cite{two loops} not only confirms this structure, but shows that reasonable
estimates for the relevant effective coupling constants lead to very sharp
predictions for the scattering lengths and effective ranges.}

In current algebra approximation,
only the I=0 component $M_0(s)$ is different
from zero. According to eqs.(1) and (2), the leading term is
given by $T(s)\equiv
(3s-4M_\pi^2)(M_\eta^2-M_\pi^2)^{-1}$.
The one loop calculation supplements this expression with
corrections of first nonleading order and also gives rise to components
with I=1,2.
We denote the one loop approximation by
$\Mbar_0(s),\Mbar_1(s),\Mbar_2(s)$.
In the notation of ref.\cite{GL eta}, the result reads
\bea\label{one loop}
\Mbar_0(s) \al=\al T(s)+\Delta_0(s)\{1+\mbox{$\frac{2}{3}$}T(s)\}
+\Delta_2(s)\{1-\mbox{$\frac{1}{3}$}T(s)\}+\Delta_3(s)+V(s)\no
\al\al+\mbox{$\frac{2}{3}$}\Delta_{\mbox{\tiny GMO}}
T(s) +\mbox{$\frac{8}{3}$}M_\pi^2\,\{
\Delta_{\mbox{\tiny F}}T(s)-\Delta_{\mbox{\tiny GMO}}\}
(M_\eta^2-M_\pi^2)^{-1}\co\no
\Mbar_1(s)\al=\al\{\mbox{$\frac{3}{2}$}\Delta_1(s)-4 s L_3
F_\pi^{-2}\} (M_\eta^2-M_\pi^2)^{-1}\co\no
\Mbar_2(s)\al=\al\mbox{$\frac{1}{2}$}\Delta_2(s)\{3-T(s)\}\fs
\nonumber\eea
The functions $\Delta_0(s),\ldots\,,\Delta_3(s)$ are
generated by the final
state interaction. They arise from one loop graphs, contain cuts
along the real axis and are inversely proportional to $F_\pi^2$.
The quantity
$V(s)$ represents a subtraction term linear in $s$, which also
originates in these graphs (for explicit expressions,
see ref.\cite{GL eta}).
The coefficients $\Delta_{\mbox{\tiny GMO}}$ and $\Delta_{\mbox{\tiny F}}$
are known low energy
constants related to the masses and decay constants of the pseudoscalar octet.
Finally, the term $L_3$ is an effective coupling constant of
the chiral Lagrangian, whose value is known from $K_{\ell 4}$-decay
\cite{Bijnens Colangelo Gasser}.

The one loop formula
shows that, up to
and including first order corrections, the decay rate is inversely proportional
to $Q^4$, with a known factor of proportionality. A measurement
of the rate thus amounts to a measurement of the quark mass ratio $Q$.
Apart from experimental uncertainties, the accuracy
of the result for $Q$ is determined by the accuracy of the theoretical
prediction for the factor $M(s,t,u)$.
This motivates the present work.

In fig.1, the
real part of the one loop amplitude is plotted along the line
$s=u$ (dash-dotted).
For small values of $s$,
the curve very closely follows the current
algebra formula (dashed) -- there, the correction is small. The cusp
at $s=4\, M_\pi^2$ arises from the final state interaction, which
is dominated by the isospin zero S-wave. The
singularity requires the amplitude to
grow more strongly with $s$ than the current algebra result.
At the maximal energy
accessible in $\eta$-decay, $\sqrt{s}=M_\eta-M_\pi$, the correction amounts to
$\frac{2}{3}$ of the leading term.
This indicates that (a) the current algebra formula underestimates the rate by
a substantial factor and (b) the first two terms of the chiral perturbation
series represent a good approximation only at small values of $s$.

The reason why in part of the physical region, the corrections are unusually
large is understood. Chiral symmetry implies that the pions are subject to
an interaction with a strength determined by $F_\pi$.
In the chiral limit the corresponding I=0 S-wave phase shift is given by
$\delta_0^0=s/16\pi^2F_\pi^2$ and thus grows
with the square of the center
of mass energy. Accordingly, the corrections rapidly
grow with the energy of the charged pion pair.

\begin{figure}

\centering
\mbox{\epsfysize=8cm \epsfbox{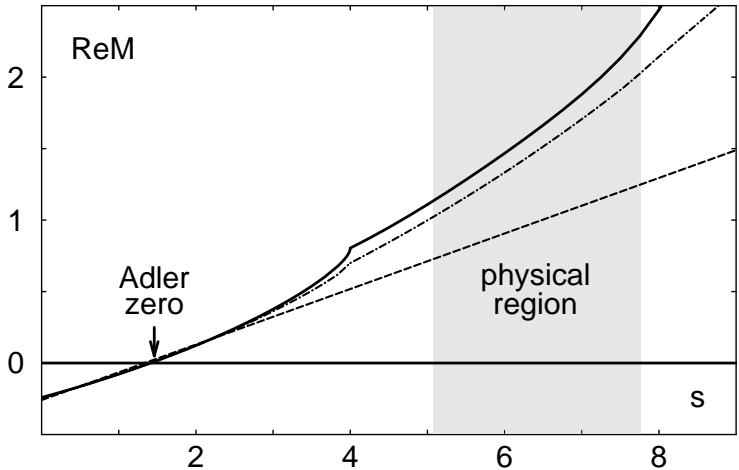} }
\vspace*{-2cm}
\caption{Real part of the amplitude along the line $s=u$.
The dashed line is the current algebra prediction. The dash-dotted curve
corresponds to the one loop approximation of chiral perturbation theory.
The full line represents
the result of the dispersive calculation described in the text.
}\end{figure}

{\it 2. Dispersion relations and sum rules}.
There is a well-known method which allows one to calculate the final state
interaction effects, also if they are large: dispersion relations. Indeed, this
method was applied to three-body decays long
ago \cite{history} and many papers
concerning the decay $\eta\!\rightarrow\!3\pi$ have appeared in the meantime --
they may be traced with refs.\cite{Neveu
Scherk}--\cite{Kambor
Wiesendanger Wyler}. Crudely speaking, analyticity and unitarity
determine
the decay amplitude up to the subtraction constants.
Chiral perturbation theory is needed only
for these. The additional information contained in the one loop result
merely reflects the fact that the chiral
perturbation series obeys unitarity, order by order. In the following,
we determine the absorptive part with unitarity and use
dispersion relations to express the full amplitude in terms of the
absorptive part and a set of subtraction constants. The latter are calculated
by matching the dispersive and one loop representations
at low energies. In principle, the same steps
must be taken whenever unitarity is used to extend the range of validity of the
effective theory -- our analysis outlines the general procedure in the context
of a nontrivial example. For a recent report on the marriage of chiral
perturbation theory and dispersion relations, see ref.\cite{Donoghue}.

We first set up the dispersion relations obeyed by the one loop amplitude.
The functions $\Mbar_0(s),\Mbar_1(s),\Mbar_2(s)$ are analytic except for a cut
along the real axis. For $s\!\rightarrow\!\infty$,
$\Mbar_0(s)$ and $\Mbar_2(s)$ grow like $s^2\,\mbox{log} s$, while
$\Mbar_1(s)$ is of order $s\,\mbox{log}s$. Accordingly, three
subtractions are needed for $\Mbar_0(s),\Mbar_2(s)$ and two for $\Mbar_1(s)$,
\bea\label{DR 1} \Mbar_0(s)\al=\al a_0+b_0 s +c_0 s^2
+\frac{s^3}{\pi}\!\int_{4M_\pi^2}^{\infty}
\frac{ds'}{s'^{\,3}}\frac{\mbox{Im}\Mbar_0(s')}{s'-s-i\epsilon}\co\nonumber\\
 \Mbar_1(s)\al=\al a_1+b_1 s
+\frac{s^2}{\pi}\!\int_{4M_\pi^2}^{\infty}
\frac{ds'}{s'^{\,2}}\frac{\mbox{Im}\Mbar_1(s')}{s'-s-i\epsilon}\co\\
\Mbar_2(s)\al=\al a_2+b_2 s +c_2 s^2
+\frac{s^3}{\pi}\!\int_{4M_\pi^2}^{\infty}
\frac{ds'}{s'^{\,3}}\frac{\mbox{Im}\Mbar_2(s')}{s'-s-i\epsilon}\fs
\nonumber\eea
Inserting this representation in eq.(\ref{isospin decomposition}), the
one loop amplitude takes
the form $M(s,t,u)=P(s,t,u)+D(s,t,u)$, where the polynomial $P(s,t,u)$ contains
the contributions from the subtraction constants $a_0,\ldots,c_2$ while
$D(s,t,u)$ collects the dispersion integrals. On kinematic grounds,
the polynomial only involves four independent terms, which may be written in
the form $P(s,t,u)=a + b\, s +c\,s^2-d\, (s^2+2\, t\, u) $: Only
four combinations of the eight subtraction constants $a_0,\ldots, c_2$ are of
physical significance.

The dispersive representation of the one loop amplitude thus appears to involve
four subtraction constants. In fact, however, the constant $c$ is
determined by the imaginary parts
through a sum rule, for the following reason. The
amplitudes $\Mbar_0(s)$ and $\Mbar_2(s)$
individually grow like $s^2\,\mbox{log}s$ for $s\!\rightarrow\!\infty$, but the
combination $\Mbar_0(s)+\frac{4}{3}\Mbar_2(s)$ only diverges in proportion to
$s\,\mbox{log}s$. For the above representation to reproduce
this behaviour at large values of $s$, the term
$\propto\! s^2$ arising from the dispersion integrals must be compensated by
the one from the subtraction polynomials, i.e.
\be\label{SR 1} c= c_0+\mbox{$\frac{4}{3}$}c_2=\frac{1}{\pi}
\int_{4M_\pi^2}^{\infty}
\frac{ds'}{s'^{\,3}}\{\mbox{Im}\Mbar_0(s')+\mbox{$\frac{4}{3}$}\,
\mbox{Im}\Mbar_2(s')\}\fs\ee
The terms of order $s^2\,\mbox{log}s$ also cancel in the combination
$s\,\Mbar_1(s)+\Mbar_2(s)$, but this function does not obey a twice subtracted
dispersion relation, because the term $\propto L_3$ gives rise to a
contribution which grows like $s^2$. Hence
$d$ is related to the effective coupling constant $L_3$,
\be\label{SR 2} d=
b_1+c_2=-\frac{(4L_3-1/64\pi^2)}{F_\pi^2(M_\eta^2-M_\pi^2)}
+\frac{1}{\pi}
\int_{4M_\pi^2}^{\infty}
\frac{ds'}{s'^{\,3}}\{s'\,\mbox{Im}\Mbar_1(s')+
\mbox{Im}\Mbar_2(s')\}\fs\ee

This shows that the one loop amplitude is fully determined by its
imaginary parts and by the three constants
$a,b$ and $L_3$. The dispersive representation explicitly expresses the
fact that an analytic function is uniquely determined by its
singularities: While the functions $\mbox{Im}\Mbar_\I(s)$
describe
those associated with the branch cut along the real axis, the
subtraction constants account for the singularities at infinity.

{\it 3. Unitarity}.
In one loop approximation, the amplitude obeys unitarity only modulo
contributions of order $p^6$. The corresponding expressions for the
discontinuities across the
branch cut only hold to leading order of the chiral expansion.
The unitarity condition may be solved more accurately, using an approximation
which holds up to and
including two loops of chiral perturbation theory. The
approximation
accounts for two-body
collisions, but disregards multiparticle interactions. In
the case of elastic scattering, the approximation is well-known and
is referred to as elastic unitarity. The extension
of this notion to three-body decays is not trivial, however: Some of the early
papers contain erroneous prescriptions.
A coherent framework is obtained with analytic continuation in the mass
of the $\eta$. One
first considers the scattering amplitude $\pi\eta\!\rightarrow\!\pi\pi$ for a
value of the strange quark mass for which $M_\eta\!<\!3M_\pi$, such that
the decay is kinematically forbidden. Elastic
unitarity then represents a consistent approximation scheme.
In this approximation, only two-pion intermediate states are
taken into account, such that the imaginary part of the scattering amplitude
coincides
with the discontinuity across the two-pion cut and is determined by the
phase shifts of $\pi\pi$ scattering. Denoting the discontinuity of the
function $\MI(s)$ by
$\mbox{disc}\,\MI(s)\!\equiv\!\{\MI(s+i\epsilon)-\MI(s-i\epsilon)\}/2i$,
the elastic unitarity conditions read\footnote{
In the standard notation, where the phase shift
of the partial wave with isospin and angular momentum quantum
numbers $\mbox{I},\ell$ is denoted by
$\delta^\I_\ell(s)$, the unitarity condition for the amplitudes
$M_0(s),M_1(s)$ and $M_2(s)$ involves $\delta_0(s)\equiv\delta_0^0(s),
\delta_1(s)\equiv \delta_1^1(s)$ and $\delta_2(s)\equiv\delta_0^2(s)$,
respectively.}
\be\label{unitarity}
\mbox{disc}\,\MI(s)=\sin\delta_\I(s)\,
e^{-i\delta_\Ism(s)}
\{\MI(s)+\MIhat(s)\}\fs\ee
The first term in the curly brackets accounts for
repeated collisions in the $s$-channel, while the second arises from
two-particle final state interactions
in the $t$- and $u$-channels and involves angular averages of the type
\bea &\!\!& \!\! \langle
z^n\MI\,\rangle (s)\equiv
\mbox{$\frac{1}{2}$}\int_{-1}^{+1}\!dz
z^n \MI\,(\mbox{$\frac{3}{2}$}s_0-
\mbox{$\frac{1}{2}$}s+\mbox{$\frac{1}{2}$}\kappa z)\co\no
&\!\!&\!\!\kappa\equiv\left\{(1-4M_\pi^2/s)
\{(M_\eta +M_\pi)^2 -s\}\{(M_\eta -M_\pi)^2 -s\}\right\}^\frac{1}{2}
\co\nonumber\eea
with $s_0\!\equiv\!\frac{1}{3}M_\eta^2+M_\pi^2$.
In this notation, the explicit
expression
for the discontinuity due to two-body collisions in the crossed channels reads
\bea \hat{M}_0&\!\!=&\!\! \mbox{$\frac{2}{3}$}\langle M_0\rangle
+\mbox{$\frac{20}{9}$} \langle M_2\rangle
+2(s-s_0)\langle
M_1\rangle+\mbox{$\frac{2}{3}$}\kappa \,\langle zM_1\rangle\no
\hat{M}_1&\!\!=&\!\!\kappa^{-1}\,\{3\langle zM_0\rangle
-5\langle z M_2\rangle
+\mbox{$\frac{9}{2}$}(s-s_0)\langle
zM_1\rangle+\mbox{$\frac{3}{2}$}\kappa\, \langle z^2M_1\rangle\}\no
\hat{M}_2&\!\!=&\!\!\langle M_0\rangle +\mbox{$\frac{1}{3}$}
\langle M_2\rangle
-\mbox{$\frac{3}{2}$}(s-s_0)\langle
M_1\rangle-\mbox{$\frac{1}{2}$}\kappa \,\langle zM_1\rangle\fs
\nonumber\eea
In contrast to the imaginary parts, the
discontinuities are analytic in $M_\eta$, so that the above relations
may be continued to the physical value of the mass.
While the
imaginary parts are real by definition, the two particle discontinuities
obtained through analytic continuation are complex. In the unitarity
relation for the decay amplitude, these contributions arise from
the disconnected part of the $T$-matrix,
describing two-particle collisions with the third pion as a spectator. The
imaginary part of the amplitude necessarily also receives contributions
from the connected part.
In the above approximation, these arise from successive collisions among two
different pairs of pions. Note also that the dispersion integrals extend
beyond the physical region of the decay, where $\kappa$ may develop an
imaginary part. The integration over $z$ must then be
deformed into the complex plane, in such a way that the path avoids the
branch cut (for details, see ref.\cite{Anisovich}).

The conditions
(\ref{unitarity}) specify the approximate form of unitarity used in our
analysis. We briefly comment on the three phase shifts $\delta_0(s),
\delta_1(s),\delta_2(s)$ which occur in these conditions and which
represent an important input of our calculation.
Chiral perturbation theory yields remarkably
sharp predictions near
threshold. Together with
the experimental information available at higher energies, the Roy equations
then determine their behaviour up to $K\Kbar$ threshold, to within
small uncertainties \cite{Buettiker}. The function $\delta_0(s)$ slowly
passes
through $\frac{1}{2}\pi$ at $\sqrt{s}=850-900$ MeV and continues rising
beyond this point. In the context of our calculation, the behaviour
at higher energies does not play a crucial role.
We use a parametrization where $\delta_0(s)$ tends to $\pi$
for $s\rightarrow\infty$, so that the discontinuity, which is proportional
to $\sin\delta_0(s)$, tends to zero.
A similar representation of the high energy behaviour is also used for
the P-wave, where the phase
shift rapidly passes through $\frac{1}{2}\pi$ at $s=M_\rho^2$. The exotic
partial wave with I=2, on the other hand, describes a repulsive interaction
and does not exhibit resonance behaviour. We use a parametrization where
the function $\delta_2(s)$ asymptotically tends to zero.

{\it 4. Ambiguities in the solution of the dispersion relations.}
If the phase shifts are known, the unitarity conditions determine the
discontinuity across the branch cut in terms of the
amplitude itself. Analyticity then implies that the functions $M_0(s),M_1(s)$
and $M_2(s)$ obey dispersion relations analogous to eq.(\ref{DR 1}),
\be\label{DR 3} \MI(s)= P_\I(s)+
\frac{1}{\pi}\int_{4M_\pi^2}^{\infty}\!ds'\,\frac{\sin \delta_\I(s')
e^{-i\delta_\Ism (s')}}
{s'-s-i\epsilon}
\{\MI(s')+\MIhat (s')\}\fs \ee
For simplicity, we assume that the phase shifts rapidly reach their
asymptotic values, so that the dispersion integrals converge without
subtractions, but
this is not essential -- we might just as well write the integrals
in subtracted form, absorbing the difference in the polynomials $P_\I(s)$.

It is crucial that the dispersion relations
used uniquely determine the amplitude. In fact for the above relations this
is not the case: The corresponding homogeneous equations -- obtained by
dropping the polynomials $P_\I(s)$ -- possess nontrivial
solutions. In its simplest form, the problem shows up if the
contributions to the discontinuity from the angular averages over
the crossed channels are dropped. Unitarity then reduces
to three independent constraints of the form
$\mbox{disc}\MI\!=\!\sin\delta_\I\, e^{-i\delta_\Ism}
\MI$, or, equivalently, $\MI(s+i\epsilon)
\!=\!e^{2i\delta_\Ism}\MI(s-i\epsilon)$. This condition
is well-known from the dispersive analysis of form factors and
can be
solved explicitly: The Omn\`{e}s function, defined by
\bdm \Omega_\I(s)=\exp\left\{\frac{s}{\pi}\int_{4M_\pi^2}^\infty
\!\frac{ds'}{s'}\frac{\delta_\I(s')}{(s'-s-i\epsilon)}\right\}\co\edm
obeys
$\Omega_\I(s+i\epsilon)\!=\!
e^{2i\delta_\Ism}\,\Omega_\I(s-i\epsilon)$, so that the ratio
$m_\I(s)\!=\!\MI(s)/\Omega_\I(s)$ is continuous
across the cut. Since $\Omega_\I(s)$ does not have any zeros,
$m_\I(s)$ is an entire function. In view of the asymptotic behaviour
of the phase shifts specified above, $\Omega_0(s),\Omega_1(s)$
tend to zero in inverse proportion to $s$, while $\Omega_2(s)$ approaches
a constant. Assuming that
$\MI(s)$ does not grow faster than a power of $s$, the same then
also holds for
$m_\I(s)$. Being entire,
$m_0(s),m_1(s)$ and $m_2(s)$ thus represent polynomials:
The general solution of the simplified unitarity conditions
is of the
form $\MI(s)\!=\!m_\I(s)\,\Omega_\I(s)$, where
$m_\I(s)$ is a polynomial.
The ambiguity
mentioned above arises because the Omn\`{e}s factors belonging to the partial
waves with I=0,1 tend to zero for $s\rightarrow \infty$. The dispersion
relation for $M_0(s)$, e.g. implies
$M_0(s)\rightarrow P_0(s)$. If $P_0(s)$ is a polynomial of degree $n$,
$m_0(s)$ is of degee $n+1$. Hence the above dispersion relation for
$M_0(s)$ admits a one parameter family of solutions. The same is true of
$M_1(s)$, while the solution of the dispersion relation for $M_2(s)$ is
unique.

The same problem also occurs if the angular averages are not discarded.
The preceding discussion points the way towards
a solution of the problem: It suffices to
replace the above integral equations with the dispersion relations obeyed
by the functions $m_\I(s)\equiv \MI(s)/\Omega_\I(s)$.
Since the corresponding discontinuities are given by
\bdm \frac{\MI(s+i\epsilon)}{\Omega_\I(s+i\epsilon)}-
\frac{\MI(s-i\epsilon)}{\Omega_\I(s-i\epsilon)}=\frac{
\MI(s+i\epsilon)e^{\!-i\delta_\Ism}\!-\!\MI(s-i\epsilon) e^{i\delta_\Ism}}{
|\Omega_\I|} =2 i\,\frac{\sin\delta_\I \MIhat}{|\Omega_\I|}\co\edm
the dispersion relations take the form
\be\label{DR 4}
 \MI(s)=\Omega_\I(s)\left\{\tilde{P}_\I(s)+\frac{1}{\pi}
 \!\int_{4M_\pi^2}^\infty
\!ds'\,\frac{\sin\delta_\I(s')\,\MIhat(s')}
{|\Omega_\I(s')|\,(s'-s-i\epsilon)}\right\}\fs\ee
In the simplified situation considered above, where the angular averages
$\MIhat(s)$ are discarded, this form of the dispersion
relations indeed unambiguously fixes the solution in terms of the
polynomials $\tilde{P}_\I(s)$. Our numerical results indicate that the same
is true also for the full set of coupled integral equations, but we do
not have an analytic proof of this statement.

In the present context, the asymptotic behaviour of the phase shifts is
an academic issue. The fact that the solution of eq.(\ref{DR 3})
is unique if all of the phase shifts tend to zero, but becomes
ambiguous if some of them tend to $\pi$,
shows that this form of the dispersive
framework is deficient.
As an illustration, we return to the simplified problem and discuss the
change in the function $M_0(s)$ which results if
the high energy behaviour of the phase shift $\delta_0(s)$ is modified.
Suppose that $\delta_0(s)$ is taken to rapidly drop to zero at some large
value $s_0$, such that the corresponding
Omn\`{e}s function differs from the original one by the factor
$1-s/s_0$. If the dispersion relations are written in the form (\ref{DR 4}),
the solution merely picks up this factor and thus only receives a small
correction at low energies. With eq.(\ref{DR 3}), however,
the modification generates a qualitative change, as it selects
a unique member from a one parameter family of solutions -- the result cannot
be trusted, because it is sensitive to physically irrelevant modifications of
the input.

{\it 5. Subtractions.}
The high energy behaviour of the amplitude is the same as
in the case of elastic scattering and is dominated
by Pomeron exchange (since the amplitude
$\pi\eta\rightarrow\pi\pi$ violates isospin symmetry, the coupling to the
Pomeron is suppressed by a factor of $m_d\!-\!m_u$, but this holds
for the entire amplitude). Comparison with $\pi\pi$-scattering indicates that,
in principle, the dispersive representation of the three functions
$M_0(s),M_1(s),M_2(s)$ should require only two
subtractions. The chiral perturbation series yields an oversubtracted
representation: The number of subtractions grows with the order at which the
series is evaluated.

If only two
subtractions are made, the contributions from the region above
$K\Kbar$-threshold
are quite important and need to be analyzed in detail
(compare e.g.
ref.\cite{DGL}, where the scalar form factors are studied by using unitarity
also for the quasi-elastic transition $\pi\pi\!\rightarrow\!K\Kbar$).
The
uncertainties associated with the high energy part of the dispersion integrals
do not limit the accuracy of the measurement of $Q$, however. One may use more
than the minimal number of subtractions and determine the extra subtraction
constants with the observed Dalitz plot distribution.
With sufficiently many subtractions,
the contributions from the high energy region become negligibly small.
In principle, a
measurement
of $Q$ requires theoretical input only for the overall normalization,
while the remaining subtraction constants, which specify the dependence of
the amplitude on $s,t,u$, may be determined experimentally.

We do not invoke the observed Dalitz plot distribution, but use chiral
perturbation theory to determine the subtraction constants.
The number of subtractions is chosen such that $M(s,t,u)$ grows
at most linearly in all directions $s,t,u\!\rightarrow\!\infty$:
$M_0(s)\!=\!O(s),M_1(s)\!=\!O(1),M_2(s)\!=\!O(s)$. In view of the Omn\`{e}s
factors, this implies that the subtraction polynomials are of the form
$\tilde{P}_0(s)\!=\!
\alpha_0\!+\!\beta_0s\!+\!\gamma_0s^2,\tilde{P}_1(s)\!=\!
\alpha_1\!+\!\beta_1 s,
\tilde{P}_2(s)\!=\!
\alpha_2\!+\!\beta_2 s$ and thus involve altogether seven subtraction constants.
As it is the case with the one loop amplitude, the decomposition into the
three isospin components $M_0(s),M_1(s),M_2(s)$ is not unique. Since the
transformation $M_1(s)\rightarrow M_1(s)+k_1$, $M_2(s)\rightarrow M_2(s)+k_2+
k_3\, s$ preserves the asymptotic behaviour and can be absorbed
in $M_0(s)$, only four of the seven subtraction
constants are of physical significance. Without loss of generality we
exploit this freedom in the
isospin decomposition and set $\alpha_1\!=\!\alpha_2\!=\!\beta_2\!=\!0$.
The dispersion relations then take the form
\bea\label{DR 5}
M_0(s)\al=\al\Omega_0(s)\left\{\alpha_0+\beta_0 s+\gamma_0 s^2+
\frac{s^2}{\pi}\!\int_{4M_\pi^2}^\infty \!\frac{ds'}{s^{\prime \,2}}
\frac{\sin\delta_0(s')\,\hat{M}_0(s')}{|\Omega_0(s')|\,
(s'-s-i\epsilon)}\right\}\no
M_1(s)\al=\al\Omega_1(s)\left\{\beta_1 s+
\frac{s}{\pi}\!\int_{4M_\pi^2}^\infty \!\frac{ds'}{s'}
\frac{\sin\delta_1(s')\,\hat{M}_1(s')}
{|\Omega_1(s')|\,(s'-s-i\epsilon)}\right\}\\
M_2(s)\al=\al\Omega_2(s)
\frac{s^2}{\pi}\!\int_{4M_\pi^2}^\infty \!\frac{ds'}{s^{\prime \,2}}
\frac{\sin\delta_2(s')\,\hat{M}_2(s')}
{|\Omega_2(s')|\,(s'-s-i\epsilon)}\fs\nonumber\eea
The four subtraction constants $\alpha_0,\beta_0,\gamma_0,\beta_1$
play the same
role as the terms $a,b,c,d$ occurring in the one loop result.

The constants $\alpha_0,\beta_0$
already appear in the current algebra approximation,
which corresponds to $\alpha_0+\beta_0\,s\!
=\!(3s-4M_\pi^2)/(M_\eta^2-M_\pi^2)$.
At leading order, the amplitude thus vanishes at
$s\!=\!\frac{4}{3}M_\pi^2$,
a feature which is beautifully confirmed by the observed Dalitz plot
distribution. The phenomenon is a consequence of
SU(2)$_\R\times$SU(2)$_\L$-symmetry
and does not rely on the expansion in powers of $m_s$, which is generating the
main uncertainties in the chiral representation:
In the limit $m_u\!=\!m_d\!=\!0$, where the pions
are massless, the decay amplitude contains two Adler zeros,
one at $p_{\pi^+}\!\!=\!0\,(s\!=\!u\!=\!0)$, the
other at $p_{\pi^-}\!\!=\!0\,(s\!=\!t\!=\!0)$. The perturbations generated by
$m_u,m_d$ produce a small imaginary part, but the line
$\mbox{Re}M(s,t,u)\!=\!0$
still passes through the vicinity of the above two points. The figure shows
that along the line $s\!=\!u$, the one loop corrections shift
the Adler zero from $1.33\, M_\pi^2$ to $\sA\!=\!1.41\, M_\pi^2$. The position
of the zero is related to the value of the constant $\alpha_0$,
while $\beta_0$
determines the slope of the curve there.
The figure also
shows that the corrections of order $m_s$ barely modify the latter:
The
slope of the one loop amplitude differs from the current algebra result,
$\beta_0=3/(M_\eta^2-M_\pi^2)$, by less than a percent.
We therefore assume chiral perturbation theory to be
reliable in the vicinity of the Adler zero and
determine the subtraction constants
$\alpha_0$ and $\beta_0$ by matching the dispersive representation of the amplitude
with the one loop representation there. More specifically, we require that the
first two terms in the Taylor series
\bdm M(s,3s_0-2s,s)=M_A+(s-s_A)S_A+O\{(s-s_A)^2\}\edm
agree with those of the one loop amplitude, where
\bdm M_A=-0.0041\, i\co\;\;\;\;S_A=
(0.1957 - 0.0519 \,i)\,M_\pi^{-2}\fs\edm

Concerning the value of $\alpha_0$, this
procedure is safe, because $\alpha_0$ is protected by the low energy
theorem
mentioned above.
The slope, on the other hand,
is not controlled by SU(2)$_\R\times$SU(2)$_\L$. The one loop result does
account for the corrections of order $m_s$, but disregards higher order
contributions.
According to a general rule of thumb, first order SU(3) breaking
effects are typically of order $25\%$ while second order contributions
are of the order of the square of this. It so happens that in the case of
the slope, the various first order corrections nearly cancel one
another. The second order effects are discussed in some detail in
ref.\cite{HL mix}, on the basis of the large $N_c$ expansion. In
particular, it is shown there that
the one loop result fully accounts for $\eta\eta^\prime$ mixing
\cite{Akhoury Leurer Pich}, except for a factor
of $\cos \theta_{\eta\eta^\prime}\!\simeq\! 0.93$. This represents
a second order correction of
typical size, consistent with the rule of thumb.

In the convention adopted above, the term $\beta_1$ determines the
slope of $M_1(s)$. More generally, $\beta_1$ is related to the convention
independent quantity
\be \label{SR 3} M^\prime_1(0)+\mbox{$\frac{1}{2}$}M^{\prime\prime}_2(0)
 =\beta_1+\frac{1}{\pi}\!\int_{4M_\pi^2}^\infty \!ds'\left
 \{\frac{\sin\delta_1(s')\hat{M}_1(s')}{s^{\prime\,2}|\Omega_1(s')|}
 +\frac{\sin\delta_2(s')\hat{M}_2(s')}{s^{\prime\,3}|\Omega_2(s')|}
 \right\}\fs\ee
In one loop approximation, the lhs coincides with $b_1+c_2$. The low
energy theorem (\ref{SR 2}) thus relates $\beta_1$
to the coupling constant $L_3$, whose value
is known rather accurately from $K_{\ell_4}$ decay,
$L_3\!=\!(-3.5\pm 1.1)\cdot 10^{-3}$ \cite{Bijnens Colangelo Gasser}.

Both for the dispersive
representation and for the one loop approximation, the
quantity $M^\prime_1(0)+\frac{1}{2}M^{\prime\prime}_2(0)$ receives
two contributions:
one from an integral
over the discontinuities of the amplitude, which accounts for
the effects generated by the low lying states, and one from
a subtraction
term $\{\beta_1\leftrightarrow
-(4L_3-1/64\pi^2)/F_\pi^2(M_\eta^2-M_\pi^2)\}$, which incorporates all
other singularities, including those at infiniy.
The main difference between the two representations is that the integral
in eq.(\ref{SR 3})
exclusively accounts for the $\pi\pi$ discontinuities, while the one
in eq.(\ref{SR 2}) includes the singularities generated by
$K\Kbar,\pi\eta$ and
$\eta\eta$ intermediate states, albeit only to one loop.
One readily
checks that the integrals over the elastic
region
$4M_\pi^2\!<\!s\!<\!(M_\eta+M_\pi)^2$ only differ by terms of order
$p^2$, which are beyond the accuracy of the one loop prediction. This is
an immediate consequence of the fact that
chiral perturbation theory is consistent with unitarity. In the elastic
region, the
dispersive calculation yields a more accurate representation of the
discontinuities than the one loop approximation. Accordingly, we use this
approximation only to estimate the remainder.
The one loop result shows that
the contributions from
$\pi\eta$ intermediate states are proportional to $M_\pi^2$ and therefore
tiny -- inelastic channels generate significant effects only
for $4M_K^2\!<\!s\!<\!\infty$. The entire contribution from this interval
to the integral in eq.(\ref{SR 2}) amounts to a shift in $\beta_1$
of $+1.7 \,\mbox{GeV}^{-4}$, too small to stick out from the uncertainties in
the term from $L_3$. We conclude that $\beta_1$ is dominated by
this term,
\bdm \beta_1\simeq
-\,\frac{\;4L_3-1/64\pi^2}{F_\pi^2\,(M_\eta^2-M_\pi^2)}
=6.5\pm 1.8\,\mbox{GeV}^{-4}\fs\edm
and take the above shift as an
estimate for the uncertainties associated with the inelastic channels.

The constant $\gamma_0$ is related to the
convention independent combination $M_0^{\prime\prime}(0)
+\frac{4}{3}M_2^{\prime\prime}(0)$ of Taylor coefficients and may be
evaluated in the same manner. In particular,
one may check that the elastic contributions to the
relevant integrals coincide within the accuracy of the one loop
representation also in this case.
The only difference is that the
low energy theorem (\ref{SR 1}) does not
contain a term analogous to $L_3$, so that the arguments outlined in the
preceding paragraph now imply
$\gamma_0\simeq 0$. The uncertainties to be attached to this result
may again be estimated with the contribution from the interval
$4M_K^2\!<\!s\!<\!\infty$ to the integral over the imaginary part
of the one loop amplitude, which
amounts to a shift in the value of $\gamma_0$ by
$+8.6\, \mbox{GeV}^{-4}$.
The net effect of the uncertainties
in the subtraction constants
$\beta_1,\gamma_0$ is small. At the center of the Dalitz plot,
the above shifts due to inelastic discontinuities increase the amplitude
by about $5\,\%$. The order
of magnitude is comparable with the shift in
$\beta_0$ due to $\eta\eta^\prime$ mixing, which acts in the opposite
direction.

{\it 6. Iterative construction of the solution.}
Since the system is linear, the
dependence of the solution
on the subtraction constants $\alpha_0,\beta_0,\gamma_0$ and $\beta_1$
is of the form $\MI (s)=\alpha_0\,
f_{\hspace{0.06em}\mbox{\scriptsize I}}(s)+\beta_0\,
g_{\hspace{0.06em}\mbox{\scriptsize I}}(s)+\gamma_0\,
h_{\hspace{0.06em}\mbox{\scriptsize I}}(s)+\beta_1\,
k_{\hspace{0.06em}\mbox{\scriptsize I}}(s)$.
The coefficients $f_0(s),f_1(s),f_2(s)$ represent the
solution for $\alpha_0\!=\!1,\beta_0\!=\!\gamma_0\!=\!\beta_1\!=\!0$ and
similarly for the other terms. The solutions may
be obtained iteratively, starting e.g. with $\MI(s)\!=\!0$.
Once the iteration has
been performed for the four fundamental solutions defined above,
the values of $\alpha_0,\beta_0$ for which the Adler condition
is met are readily
worked out: Denote the total amplitudes which corresponds to the four sets of
isospin coefficients by $f(s,t,u),g(s,t,u),h(s,t,u),k(s,t,u)$, restrict these to the
line $s\!=\!u$ and calculate the values as well as the first derivatives
at the point $\sA\!=\!1.41\,M_\pi^2$. The requirement that the sum
$M=\alpha_0\, f+\beta_0\, g+\gamma_0\,h+\beta_1 k$ agrees with the one
loop amplitude there, both in value and slope, amounts to two linear
equations for $\alpha_0$ and $\beta_0$.

The numerical results obtained with the above dispersive machinery are
illustrated in fig.1:
The solid
line shows the real part of the amplitude along the line $s\!=\!u$, which
connects one of the Adler zeros with the center of the Dalitz plot. As can be
seen
from this figure, the corrections to the one loop result are rather modest.
The calculation confirms the crude estimates
given in ref.\cite{GL eta}. The error bars to be attached to the
dispersive evaluation are remarkably small. They are dominated by the
uncertainties in the low energy theorems used to determine the
subtraction constants. A more detailed account of our numerical results
is in preparation \cite{AL results}.

We are indebted to V. V. Anisovich, Hans Bebie, J\"{u}rg Gasser and Peter
Minkowski for useful comments and generous help with computer ethono\-graphy
and to Joachim Kambor, Christian Wiesendanger and Daniel Wyler
for keeping us informed about
closely related work \cite{Kambor Wiesendanger Wyler}.

\end{document}